\documentclass[useAMS,fleqn,usenatbib]{mnras}
\usepackage[T1]{fontenc}
\usepackage{ae,aecompl}

\usepackage{graphicx}	
\usepackage{amsmath}	
\usepackage{amssymb}	

\usepackage{enumerate}
\usepackage{color}
\usepackage{graphicx}
\usepackage{multicol}
\usepackage{savesym}
\usepackage{color}
\usepackage{tipa}

\newcommand{\OVII}{\mbox{{O\,{\sevensize VII}}}}

\newcommand{\Ka}{\ensuremath{\hbox{K}\alpha~}}
\newcommand{\Kb}{\ensuremath{\hbox{K}\beta~}}

\def\xmm{{\it XMM-Newton }}
\def\cha{{\it Chandra }}

\def\nus{{\it NuSTAR }}

\title[Pulsating Fe K$\alpha$ line in GX 301-2]{Discovery of a pulsating Fe K$\alpha$ line in GX 301-2} 
\author[J. Liu et al.]{Jiren Liu$^{1}$\thanks{E-mail: jirenliu@nao.cas.cn}, Roberto Soria$^{2,3,4}$, Erlin Qiao$^{1,2}$, and Jifeng Liu$^{1,2}$\\
	 $^{1}$National Astronomical Observatories, 20A Datun Road, Beijing 100012, China\\
$^{2}$College of Astronomy and Space Sciences, University of Chinese Academy of Sciences, Beijing 100049, China \\
$^{3}$International Centre for Radio Astronomy Research, Curtin University, GPO Box U1987, Perth, WA 6845, Australia\\
$^{4}$Sydney Institute for Astronomy, School of Physics A28, The University of Sydney, Sydney, NSW 2006, Australia\\
}

\date{Accepted XXX. Received YYY; in original form ZZZ}

\pubyear{2018}

\begin{document}
\label{firstpage}
\pagerange{\pageref{firstpage}--\pageref{lastpage}}
\maketitle

\begin{abstract}
We report on the detection of a pulsating Fe \Ka line in the High Mass X-ray Binary (HMXB)
GX 301-2, from a 40-ks \cha observation near periastron. The pulsations in the Fe \Ka emission appeared only in the first 7 ks of the
observation, with a period and phase profile similar to those of the continuum. The presence of pulsed fluorescent lines is an unusual property in HMXBs.
After 7 ks, the continuum flux increased by a factor of three, the Fe \Ka flux increased only by about 10\%, and the pulsating signal in the line disappeared.
Finally, in the second half of the observation, both the continuum and the line flux dropped 
by a similar factor of 2. We suggest that the pulsating component of the Fe \Ka line is 
coming from a transient non-isotropic distribution of dense gas around the neutron star, 
for example an accretion stream induced by periastron passage, or from the illuminated surface 
of the donor star.

\end{abstract}


\section{Introduction}
The classical High Mass X-ray Binary (HMXB) system GX 301-2 consists of
a pulsar immersed in a slow ($\approx$300 km\,s$^{-1}$) and dense stellar wind 
($\dot{M}_{\rm w} \sim 10^{-5} M_\odot$\,yr$^{-1}$), arising from the 
B1 hypergiant Wray 977 \citep{Kap06}. The donor star has an estimated mass of 
$\approx$39--53 $M_\odot$, a radius of $\approx$62 $R_\odot$, and a distance from us of $\approx$3 kpc \citep{Kap06}.
The orbit of the pulsar has a period $P_{\rm orb} = 41.506 \pm 0.003$ d, with reference phase 0 at the Modified Julian Date (MJD) $T_0 = 43,906.06  \pm 0.11$, and an orbital period derivative estimated as $\dot{P}_{\rm orb} = (-3.7 \pm 0.5) \times 10^{-6}$ s s$^{-1}$ \citep{Dor10}.
The system has an eccentricity $e \approx 0.46$, with a periastron distance of $\approx$25 $R_{\odot}$ between the neutron star and the surface of the donor star \citep{Sat86, Koh97}. The inclination angle is moderately high, between 44$^{\circ}$--78$^{\circ}$ \citep{LK08,Kap06}.
The pulsar has a long spin period $P_{\rm s} \approx 680$ s, which 
showed erratic changes over the last 30 years \citep[e.g.][]{Dor10,Eva10}. 
Such a long pulse period has been attributed to a strong magnetic field 
\citep[$\sim10^{14}$ G,][]{Dor10}. The field measured from the cyclotron line is much weaker ($\approx$5 $\times
10^{12}$ G), but the two values can be consistent with each other if the cyclotron line region comes from a tall
accretion column of height $\approx$2.5--3 $R_{NS}$ \citep[][but also see \citet{IF12}]{Dor10}.

\begin{figure*}
\hspace{-1.3cm}
\includegraphics[width=7.3in]{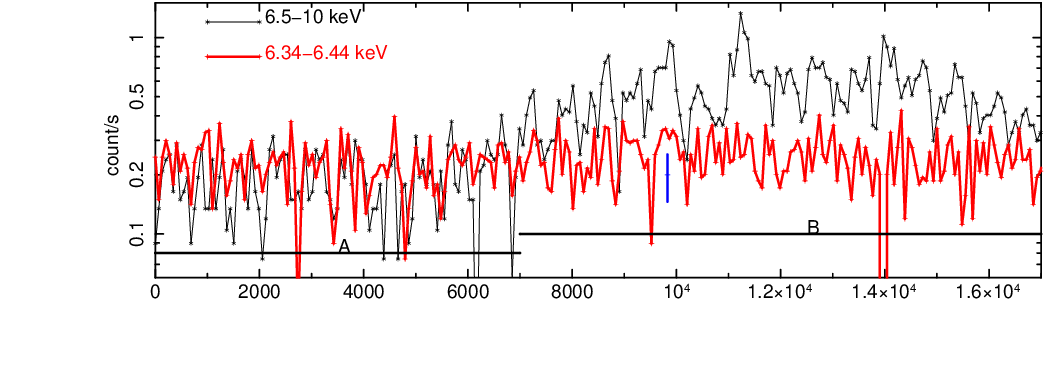}\\[-20pt]
\hspace{-1.2cm}
\includegraphics[width=7.3in]{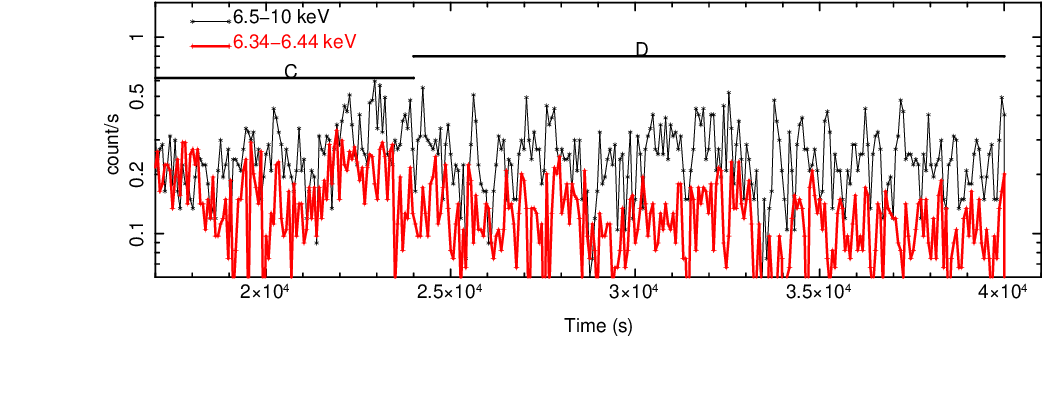}
\vspace{-1.3cm}
\caption{Light curves of the continuum-subtracted Fe \Ka line (6.34--6.44 keV) and the continuum (6.5--10 keV), for
	the first 17 ks of the observation (top panel) and for the interval between 17--40 ks (bottom panel). The time
	resolution of the data points is 68 s = 0.1 times the pulse period. Letters indicate different sub-intervals of the
	observation, used for our timing and spectral analysis. The error bar plotted in the top panel at about $10^4$ s
	represents a typical 1-$\sigma$ error, for a count rate of 0.2 ct s$^{-1}$. We took half of the count rate within the 5.8--6 keV band as the continuum level for the Fe K$\alpha$ line. We did not use the 7.0--7.2 keV band for the continuum estimate because it is contaminated by the Fe \Kb line.
}
\end{figure*}

The X-ray continuum flux of GX 301-2
shows regular orbital modulations and reaches a maximum 
about 1.4 days before periastron passage \citep{Sat86}. 
The flux increase has been explained by a gas stream induced by the neutron star near periastron 
\citep[][and references therein]{LK08, IP14}. 

GX 301-2 shows a strong 6.4-keV Fe \Ka fluorescent line, as well as many other fluorescent lines \citep{Sar96,Fur11,Suc12}.
The Fe \Ka fluorescent line is produced when the X-ray emission illuminates the surrounding
gas; its intensity and spectral profile are important tools for the study of HMXBs \citep[][and reference therein]{Tor10,Gim15}. For example, from the detection of a spectrally resolved Fe \Ka Compton shoulder at periastron, \citet{Wat03} inferred an absorption column density of $\approx$10$^{24}$ cm$^{-2}$ and an electron temperature of $\approx$0.5 eV.
The location of
the Fe \Ka fluorescent region for GX 301-2 is still actively debated. Using spectral data from the {\it Advanced Satellite for Cosmology and Astrophysics} ({\it ASCA}), 
\citet{End02} measured a Gaussian $\sigma$ of $\approx$40--80 eV for Fe \Ka and inferred an emission region within
$\approx$10$^{10}$ cm (0.3 lt-s) of the neutron star. An {\it XMM-Newton} study \citep{Fur11} did not reveal any significant time delays between the observed continuum and the reprocessed line emission, which implies a distance smaller than $\approx$6 $\times 10^{10}$ cm between the two sources of emission.
Instead, using {\it Suzaku} observations, \citet{Suc12} inferred
a distance greater than $\approx$2 $\times 10^{13}$ cm (700 lt-s), based on the relatively flat phase profile of the line
compared with that of the continuum.
Despite this discrepancy in the proposed size of the line emission region, one thing that appeared well ascertained was that the Fe \Ka line was less pulsed compared to the continuum \citep[e.g.][and reference therein]{Fur11}.
In this paper, based on {\it Chandra X-ray Observatory} observations, we show that this is not always true: the Fe \Ka line pulsates at times as strongly as the continuum.

\section{X-ray Observations}

GX 301-2 has been observed with the High Energy Transmission Grating Spectrometers (HETGs) on board {\it Chandra} on three separate occasions between 2000 and 2002. One of those observations occurred around periastron passage (ObsID 2733, 2002 January 13, exposure time of 40 ks, start time at MJD 52,287.375): this is the dataset we are reporting on for this paper. More specifically, using the ephemeris and period derivative of \cite{Dor10}, the start of the \cha observation is at orbital phase $\phi = 0.998 \pm 0.003$; the duration of the \cha observation is $\Delta \phi = 0.011$.
Spectral analysis of this dataset has been reported elsewhere \citep{Wat03}, but no timing analysis had been done.

\begin{figure}
\hspace{-0.5cm}
\includegraphics[width=2.5in]{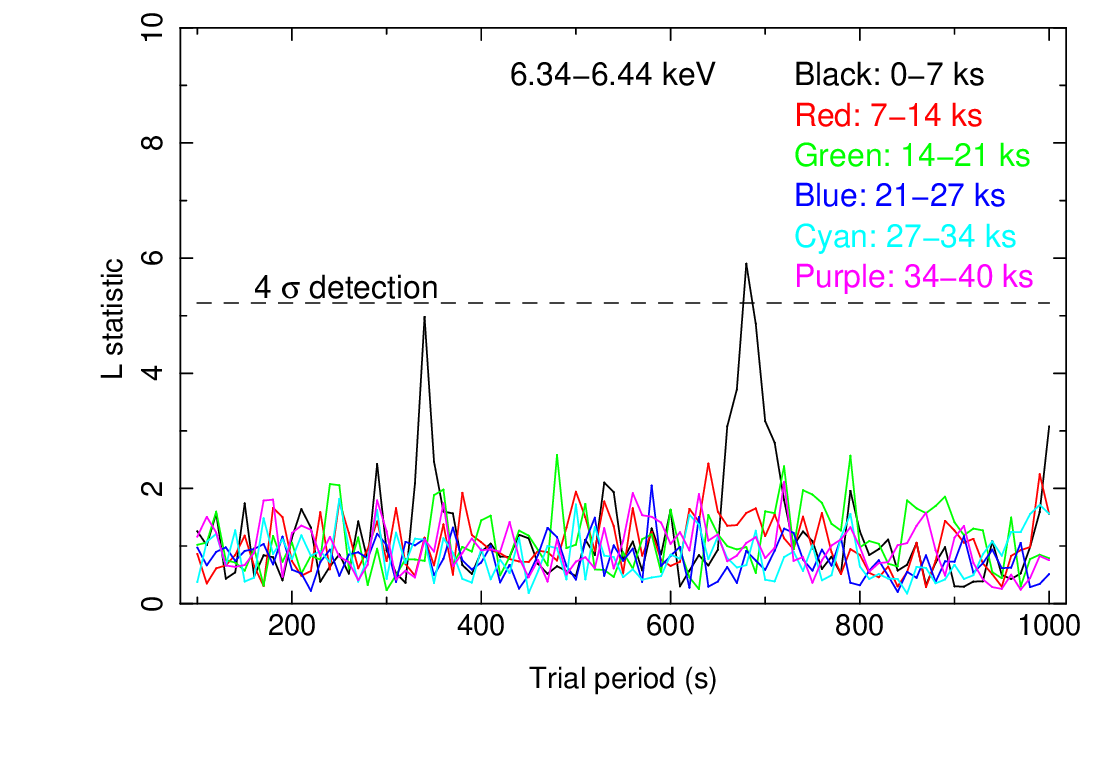}\\[-20pt]

\hspace{-0.7cm}
\includegraphics[width=2.5in]{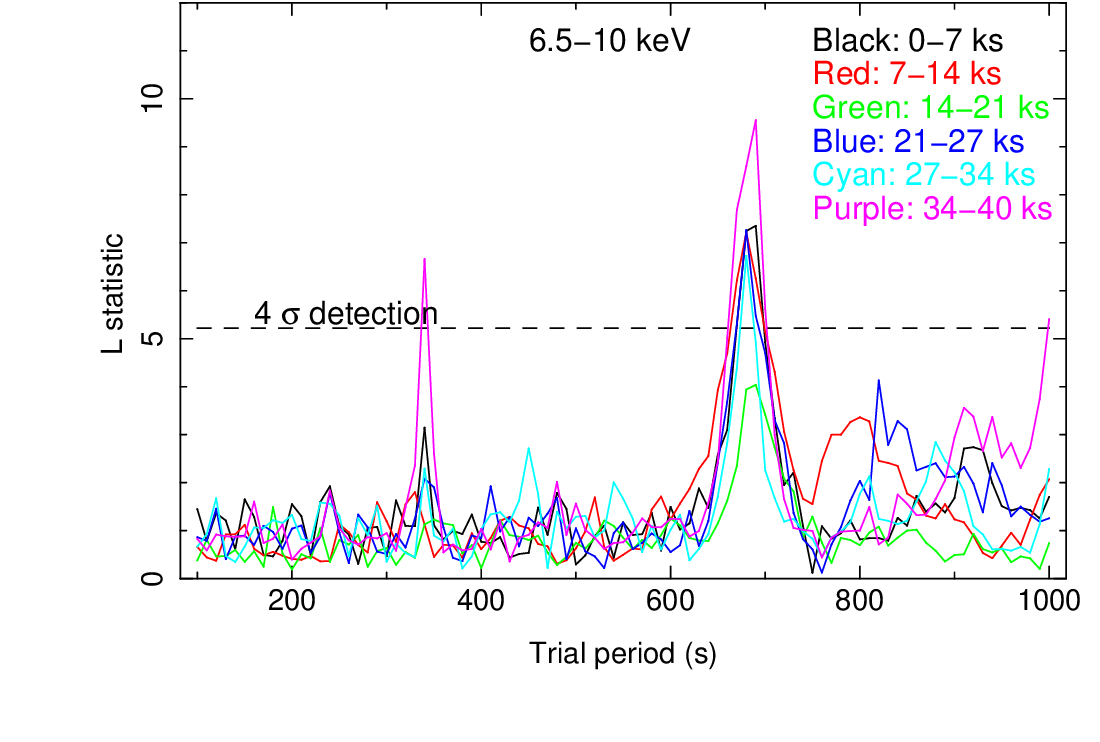}\\[-20pt]
\vspace{-0.3cm}
\caption{
Top panel: statistical significance of the pulsation period for the Fe \Ka line during different sub-intervals of the
{\it Chandra} observation; the pulse period is detected only in the first 7 ks. 
Bottom panel: statistical significance of the pulsation in the continuum. 
The dashed line shows 4 $\sigma$ detection limit.
}
\end{figure}

{\it Chandra}'s HETGs consist of two sets of gratings: the Medium Energy Grating (MEG) and the High Energy Grating (HEG).
We only used HEG data for this work, because of its better spectral resolution and larger collecting area in the Fe \Ka band. 
We downloaded the data from the public archives, and reprocessed them with the Chandra Interactive Analysis of Observations ({\sc ciao}) software version 4.8 \citep{Fru06}. We applied a barycenter correction with the {\sc ciao} tool {\it axbary}.
We extracted the HEG source spectra from a 2$''$ region in the cross-dispersion
direction with the {\sc tgcat} script \citep{TG}.

In addition, we used one archival observation of GX 301-2 taken with the {\it Nuclear Spectroscopic Telescope Array}
({\it NuSTAR}) on 2015 October 4 (ObsID 30101042002).
The system was observed at a binary phase $\approx$0.9, relatively close to periastron.
We extracted the source spectrum from a circle of $1'.5$ radius, and the background spectrum from a
blank $4'$ circle, using the {\it NuSTAR} Data Analysis Software
({\sc NuSTARDAS}\footnote{https://heasarc.gsfc.nasa.gov/docs/nustar/analysis/}). 

We used the {\it sitar\_epfold\_rate} and {\it sitar\_pfold\_rate} tools within the {\sc
sitar}\footnote{http://space.mit.edu/cxc/analysis/SITAR/distrib.html} 
software package for timing analysis. We used the Interactive Spectral Interpretation System
{\sc isis}\footnote{http://space.mit.edu/cxc/isis/} \citep{ISIS} for spectral analysis of both the {\it Chandra} and {\it NuSTAR} data.

\section{Main Results}

\subsection{Timing results}

To investigate the physical changes in the system, we divided the 40-ks observation into four sub-intervals: 0--7 ks (henceforth labelled as interval A), 7--17 ks (B), 17--24 ks (C), and 
24--40 ks (D), based on the flux levels of the Fe \Ka line (6.34--6.44 keV) and the continuum (6.5--10 keV). Half of the count rate in the 5.8--6 keV band was used for baseline continuum subtraction from the line flux. 
The light curves for the continuum-subtracted Fe \Ka line and the continuum are plotted 
in Figure 1, with a time resolution of 68 s $=$ 0.1 times the pulse period. 

In the first 7 ks of the observation, the fluxes from both the line and the continuum do not show any long-term trends, but both light curves are characterized by irregular flaring on timescales smaller than the pulse period (Figure 1). After 7 ks, in interval B, the continuum level increases; the average count rate is a factor of 3 higher than in interval A. This mini-flare is not to be confused with the so-called pre-periastron X-ray flare, which peaks at $\phi \approx 0.90$--0.95 with an amplitude 10--20 times 
that of the quiet epochs \citep[e.g.][]{IP14} and is one of the hallmarks of this HMXB. 
The average flux of the Fe \Ka line increases only by $\approx$10\% between intervals A and B. In intervals C and D, both the continuum and the line flux decreased by similar amounts. Near the end of the observation, the continuum had settled to a level slightly higher than in interval A, while the Fe \Ka line was a factor of 2 fainter than in interval A.

We searched for periodicities in the Fe \Ka line and the continuum, using the 
epoch folding method between 100--1000 s with a step-size of 10 s and 10 phase bins
for light curve (binned in 10 s) divided into 6.8 ks intervals. 
The L-statistic as defined in \citet{Dav90} is plotted in Figure 2. It obeys
an F distribution with M-1 and N-M degrees of freedom, for N data points binned in M phase bins.
The continuum shows the expected pulse periodicity 
(680 s) during the whole observation, while the Fe \Ka line shows significant 
periodicity (also at 680 s) only 
during interval A. A second peak at half of the pulsation period (340 s) is also 
significantly found for the line flux in interval A. 
The 4 $\sigma$ detection limit (P($F\leq5.22$)=99.99993\%, with 9 and 670 degrees of freedom, 
corrected for 91 searched periods) is shown as dashed line in Figure 2..

We folded the phase profile of line and continuum at different time intervals with 10 phase 
bins (Figure 3).  
For clarity the profiles of interval C (similar to those of interval D) are not plotted.
We found that the phase profiles of Fe \Ka line and continuum look similar during interval A, with broad plateaux. For
intervals B and D when no periodic pulsations of Fe \Ka line are detected, the folded phase profile of 
Fe \Ka line looks flat, as expected.
The phase profiles of the continuum during intervals B and D show a main peak, similar to those reported in the
literature \citep[e.g.][]{Fur11}. Most importantly, the pulsating behaviour of the  Fe \Ka line during interval A is a new discovery, not seen in previous observations.

The pulsed fractions defined as $f=\frac{max(p)-min(p)}{max(p)+min(p)}$ are listed in Table 1.
The pulsed fraction of the Fe \Ka line of interval A is larger than those of other 
intervals and is similar to those of the continuum.
We simulated the pulsed fractions arising from intrinsic noise of data assuming a uniform phase profile
with a Fe \Ka rate as the observed mean rate in the four intervals. The results for $10^4$ runs 
are listed in Table 1. The fake pulsed fractions are similar to those of interval B, C,
and D, but smaller than that of interval A, as expected.

\begin{figure}
\hspace{-0.7cm}
\includegraphics[width=3.6in]{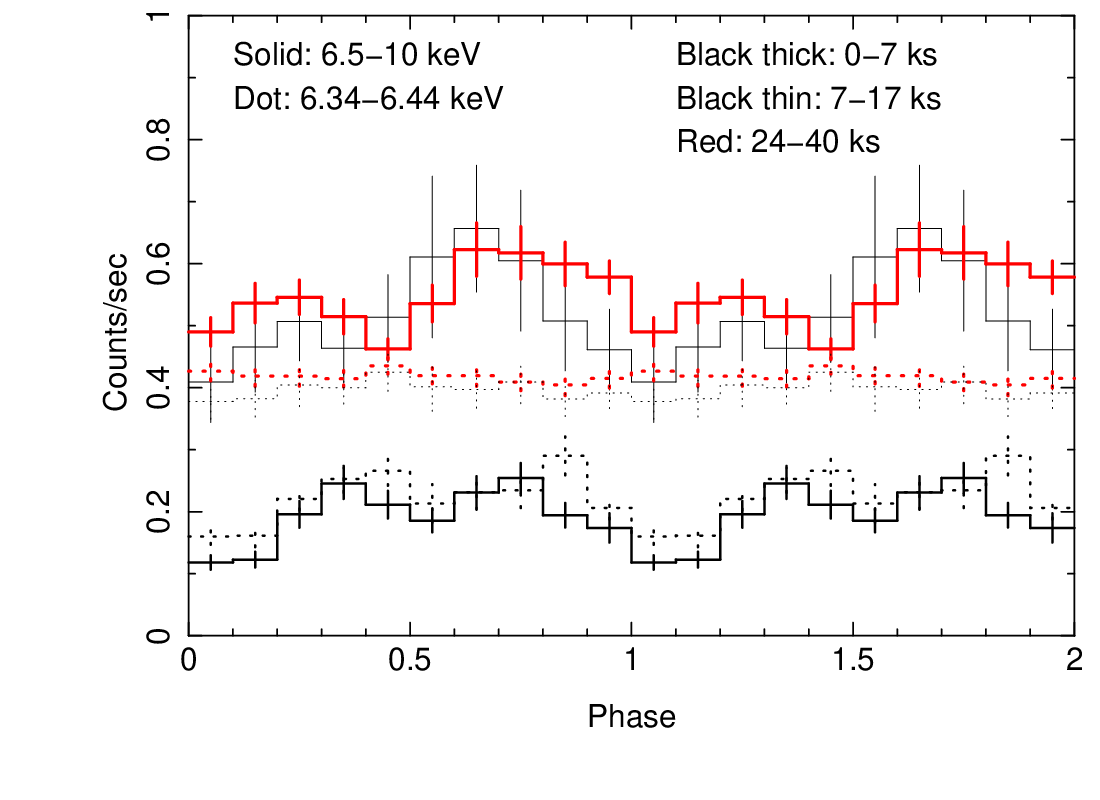}\\[-25pt]
\caption{
Phase profiles of the Fe \Ka line and the continuum for sub-intervals A (0--7 ks), B (7--17 ks), and 
D (24--40 ks).
For clarity, the pulse profiles of the Fe \Ka line for intervals B and D have been up-shifted by
0.15 and 0.3, respectively; the continuum profile of interval D has also been up-shifted by 0.3.
}
\end{figure}

\begin{figure}
\hspace{-0.7cm}
\includegraphics[width=3.6in]{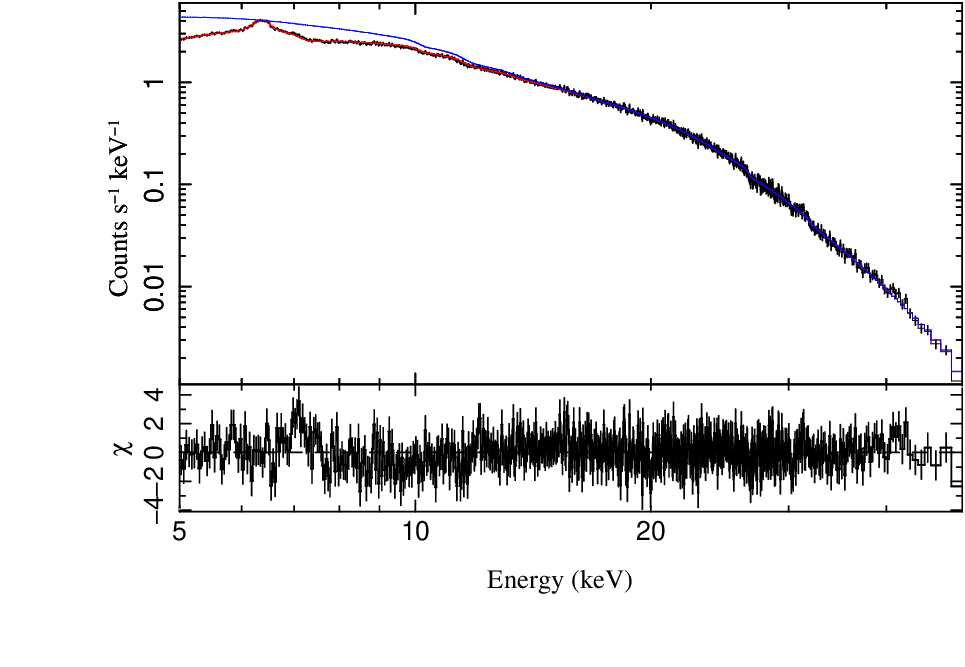}\\[-25pt]
\caption{
\nus FPMA (focal plane module
A) spectrum of GX 301-2 re-binned with a minimum S/N of 10. The blue line shows 
an unabsorbed power-law with a high-energy cutoff fitted to energies between 15 and 50 keV.
The red line shows an absorbed power-law fitted to energies between 5 and 50 keV, plus a Gaussian line.
The residuals of the absorbed power-law plus Gaussian model  are plotted in the bottom panel.
}
\end{figure}

\begin{figure}

\hspace{-0.5cm}
\includegraphics[width=3.4in]{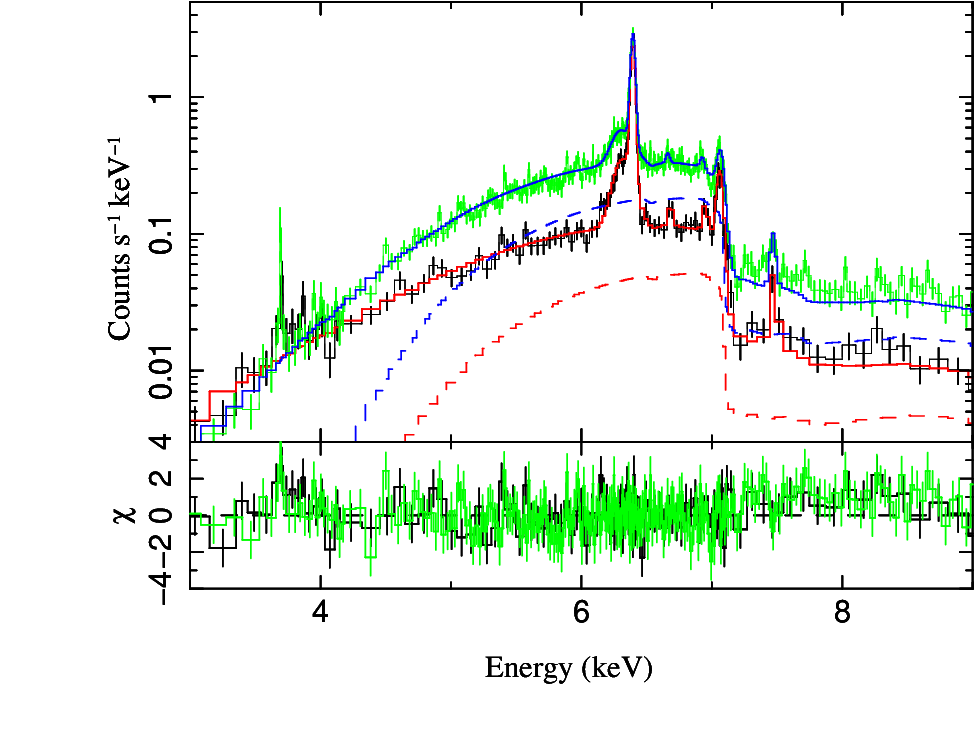}\\[-20pt]

\hspace{-0.5cm}
\includegraphics[width=3.4in]{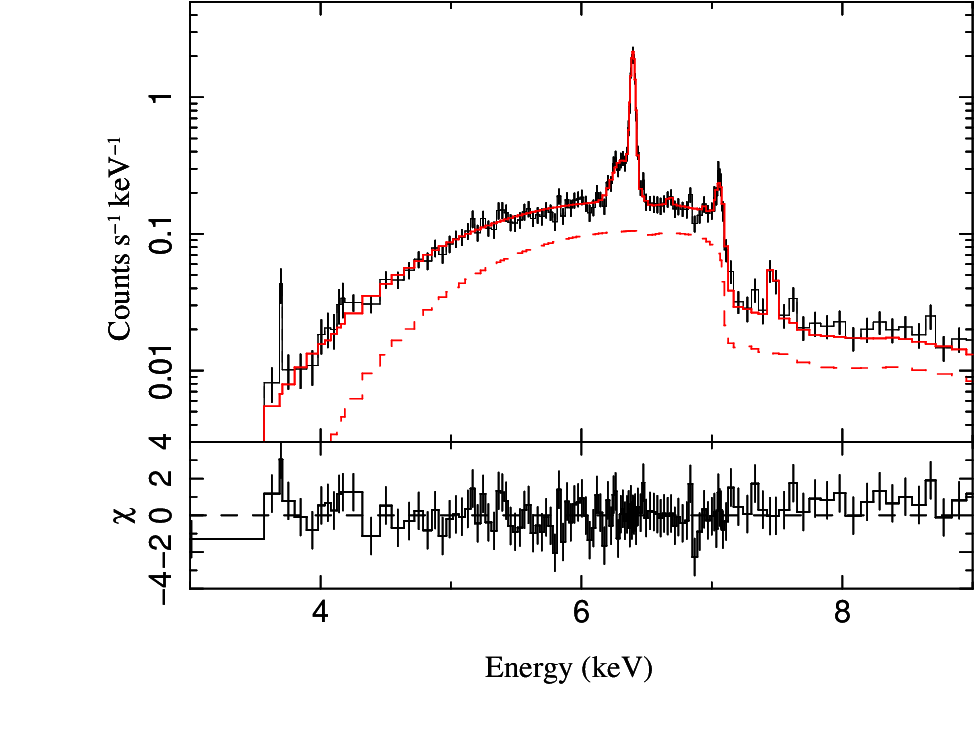}\\[-20pt]

\hspace{-0.5cm}\includegraphics[width=3.4in]{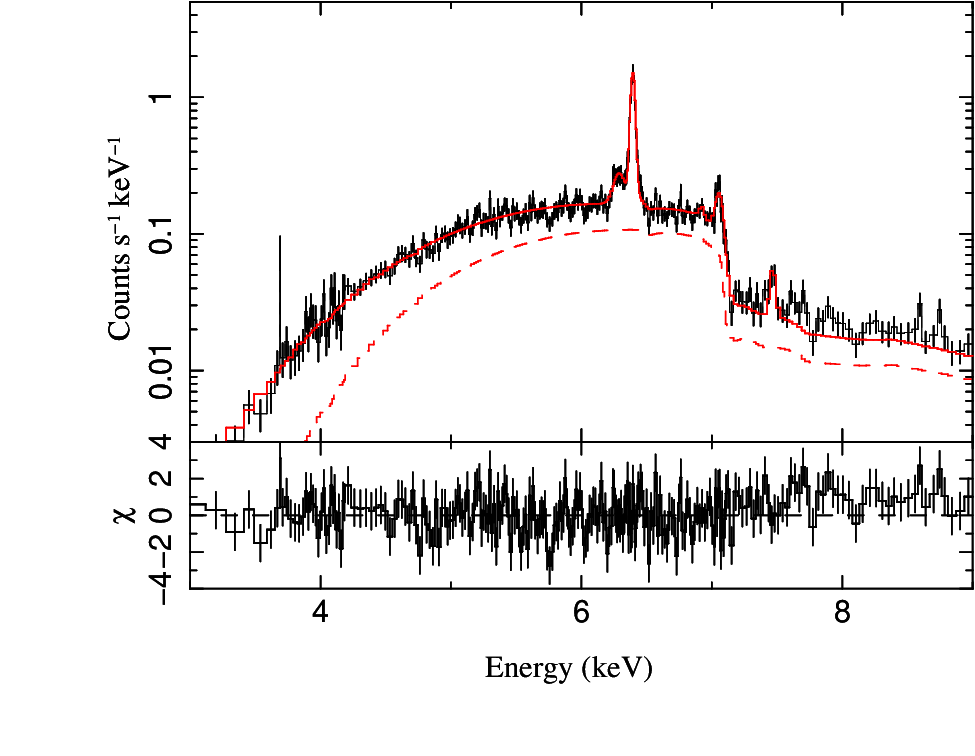}\\[-20pt]
\vspace{-0.3cm}
\caption{Top panel: \cha HEG spectrum in interval A (black data points
and residuals) and B (green data points and residuals). Both
spectra are fitted with a clumpy torus model plus transmitted intrinsic
continuum and a set of 7 Gaussian lines. The model for the interval-A
spectrum is over-plotted as a solid red histogram, while the model for
interval B is a solid blue histogram. The transmitted intrinsic
components of both models are also plotted as dashed histograms (red for
interval A and blue for interval B). The residuals $\chi$ are defined 
as the difference between data and model, divided by the data error. 
See Table 2 for the best-fitting spectral parameters and errors.
Middle panel: \cha HEG spectrum and model residuals for interval C. The
model (solid red histogram) is the same as for interval A. The dashed red
histogram shows the transmitted continuum.
Bottom panel: as in the middle panel, but for interval D.
}
\end{figure}

\begin{table}
\caption{Pulsed fraction for the four observation intervals}
\tabcolsep 1.5pt
\begin{tabular}{lccccc}
\hline
f &  Interval A &  Interval B& Interval C & Interval D  \\
\hline
Fe \Ka    &0.29$\pm$0.06&0.09$\pm$0.05&0.13$\pm$0.07&0.13$\pm$0.06 \\
Continuum    &0.37$\pm$0.06&0.23$\pm$0.04&0.24$\pm$0.06&0.33$\pm$0.04 \\
Faked Fe \Ka    &0.12$\pm$0.03&0.12$\pm$0.03&0.14$\pm$0.04&0.17$\pm$0.04 \\
\hline
\end{tabular}
\end{table}

\subsection{Spectral results}

A study of the spectral properties of the same \cha HEG observation was reported by \citet{Wat03}; however, the
observation was divided simply into two halves of 20 ks each. Thus, the unusual pulsating behaviour of the Fe \Ka line in the first 7 ks, and its subsequent sudden disappearance, went unnoticed. 
We analyzed the spectra of the four intervals A, B, C, and D separately (Figure 5). In all four intervals (but especially in interval A), the spectrum showed a strong Compton shoulder red-ward of the narrow Fe \Ka line, which implies a non-negligible contribution of the scattered component to the continuum. 

Recent studies \citep[e.g.][]{Mar17} have shown 
that stellar winds from massive stars are composed of inhomogeneous clumps.
The X-ray spectrum of a source seen through a clumpy medium is very different 
from that of a source seen through a smooth one \citep{Liu14}. For a clumpy distribution, there are many lower-density gaps between the clouds, in which low-energy X-ray photons (1--5 keV) can pass without being heavily absorbed. The clumpier the gas, the stronger the scattered low-energy emission.
This is likely why the observed low-energy spectra of GX 301-2 have generally been modelled as a partial-covering
power-law \citep[e.g.][]{Fur11}. It is a simple phenomenological model to reproduce the scattered continuum component from a clumpy material; instead, a smooth absorbing medium with the same column density ($\approx$10$^{24}$ cm$^{-2}$) would completely absorb photons below $\approx$3 keV.

\begin{table*}
\caption{Best-fitting parameters for our model (Ctorus $+$ seven Gaussian lines) in the four sub-intervals of the {\it Chandra} observation. Errors are 90\% confidence levels for one interesting parameter.}
\tabcolsep 1.5pt
\begin{tabular}{lcccc}
\hline
Parameter& Interval A &  Interval B& Interval C & Interval D  \\
\hline
\multicolumn{5}{c}{Ctorus}\\
\hline
$K_{\rm pl} \left({\mathrm{photons~keV}}^{-1} {\mathrm{~cm}}^{-2} {\mathrm{~s}}^{-1} {\mathrm{~at~1~ keV}}\right)$                       &3.8$\pm$0.6 & 7.9$\pm$2.0 & 2.4$\pm$0.7 & 1.8$\pm0.3$\\
$N^a$                         &3.5$\pm$0.6 & 5.8$\pm$0.7 & $>6.8$ & $>6.8$\\
$N^{\rm T}_{\rm {H}}$ $\left(10^{24} {\mathrm{~cm}}^{-2}\right)$    &1.5$\pm$0.5 & 1.3$\pm$0.1 & 1.0$\pm$0.1 & 0.9$\pm$0.1\\
$N^{\rm R}_{\rm {H}}$ $\left(10^{24} {\mathrm{~cm}}^{-2}\right)$    &1.6$\pm$0.3 & 1.5$\pm$0.1 & 0.9$\pm$0.1 & 0.8$\pm$0.1\\
\hline
\multicolumn{5}{c}{Neutral Fe \Ka doublet}\\
\hline
$E_1$ (keV)   & \multicolumn{4}{c}{[6.404]} \\
$E_2$ (keV)   & \multicolumn{4}{c}{[6.391]} \\
$\Delta v_{\rm{K}\alpha}$ (km s$^{-1}$)      & 180$\pm$60      & 240$\pm$30      & 210$\pm$60     & 240$\pm$60       \\
$\sigma_1 = \sigma_2$ (eV)             & 4.8$\pm$3.1  & 8.7$\pm$1.8  & 7.0$\pm$2.2 & 7.9$\pm$2.0       \\
$N_{1} + N_2$  $\left(10^{-3} {\mathrm{~photons~cm}}^{-1} {\mathrm{~s}}^{-1}\right)$               & 10.0$\pm$0.6 & 11.2$\pm$0.6 & 8.2$\pm$0.7 & 5.8$\pm$0.3      \\
EW (keV) & 0.52$\pm$0.03 &0.28$\pm$0.02 &0.38$\pm$0.03  &0.36$\pm$0.02  \\ 
\hline
\multicolumn{5}{c}{Neutral Fe \Kb}\\
\hline
$E_3$ (keV)   & \multicolumn{4}{c}{[7.058]} \\
$\Delta v_{\rm{K}\beta}$ (km s$^{-1}$) 	& -180$\pm$180     & -60$\pm$30     & -120$\pm$330   & -150$\pm$240        \\
$\sigma_3$ (eV) 	& 19$\pm$8     & 23$\pm$5     & 24$\pm$13   & 26$\pm$8        \\
$N_3$   $\left(10^{-3} {\mathrm{~photons~cm}}^{-1} {\mathrm{~s}}^{-1}\right)$      						& 2.6$\pm$0.4  & 3.1$\pm$0.4  & 2.0$\pm$0.4 & 1.6$\pm$0.3        \\
\hline
\multicolumn{5}{c}{Neutral Fe \Ka Compton shoulder}\\
\hline
$E_4$ (keV)      				& 6.32$\pm$0.01  & 6.30$\pm$0.02 & 6.31$\pm$0.03 & 6.29$\pm$0.01  \\
$\sigma_4$ (eV)    					& 63$\pm$10    & 63$\pm$13     & 60$\pm$18   & 37$\pm$12        \\
$N_4$  $\left(10^{-3} {\mathrm{~photons~cm}}^{-1} {\mathrm{~s}}^{-1}\right)$       						& 3.5$\pm$0.6  & 3.6$\pm$0.6  & 2.4$\pm$0.9 & 1.0$\pm$0.3        \\
\hline
\multicolumn{5}{c}{He-like Fe \Ka }\\
\hline
$E_5$ (keV)      				& 6.68$\pm$0.01  & 6.66$\pm$0.02 & 6.68$\pm$0.05 & 6.69$\pm$0.10  \\
$\sigma_5$ (eV)   & \multicolumn{4}{c}{[1]} \\
$N_5$  $\left(10^{-3} {\mathrm{~photons~cm}}^{-1} {\mathrm{~s}}^{-1}\right)$         						& 0.3$\pm$0.2  & 0.3$\pm$0.2  & 0.1$\pm$0.2 & 0.0$\pm$0.1        \\ 
\hline
\multicolumn{5}{c}{H-like Fe \Ka }\\
\hline
$E_6$ (keV) 						& 6.95$\pm$0.01  & 6.93$\pm$0.02 & 6.95$\pm$0.05 & 6.93$\pm$0.03  \\
$\sigma_6$ (eV)   & \multicolumn{4}{c}{[1]} \\
$N_6$   $\left(10^{-3} {\mathrm{~photons~cm}}^{-1} {\mathrm{~s}}^{-1}\right)$       						& 0.4$\pm$0.2  & 0.4$\pm$0.2  & 0.1$\pm$0.3 & 0.2$\pm$0.1        \\
\hline
\multicolumn{5}{c}{Neutral Ni \Ka }\\
\hline
$E_7$ (keV)      				& 7.47$\pm$0.02  & 7.47$\pm$0.01 & 7.47$\pm$0.01 & 7.47$\pm$0.01  \\
$\sigma_7$ (eV)   & \multicolumn{4}{c}{[1]} \\
$N_7$     $\left(10^{-3} {\mathrm{~photons~cm}}^{-1} {\mathrm{~s}}^{-1}\right)$       						& 0.7$\pm$0.3  & 1.0$\pm$0.3  & 0.8$\pm$0.3 & 0.5$\pm$0.2        \\
\hline
$\chi^2_{\nu} \left(\chi^2/{\mathrm{dof}}\right)$ &1.17 (112/96)    &1.26 (301/238)   &0.93 (98/105)   &1.31 (305/233)  \\ 
$L_{2-10}^{b}$ $\left(10^{37} {\mathrm{erg~s}}^{-1}\right)$ & 3.1$\pm$0.5 &6.4$\pm$1.6
&1.9$\pm$0.6 &1.5$\pm$0.3 \\ 
        \hline
\end{tabular}
\begin{description}
\begin{footnotesize}
\item
 Notes: $^a$: clumpiness parameter (average number of clouds along any line of sight);
 $^b$: 2--10 keV de-absorbed luminosity of the intrinsic power-law component; $N_i$ is the 
 normalization for corresponding Gaussian line.
\end{footnotesize}
\end{description}
\end{table*}

While many HMXBs show photoionization lines, the spectrum of GX 301-2 is quite different. It is
highly attenuated and dominated by neutral-like fluorescent lines \citep[e.g.][]{Fur11}. Such a
reprocessing by neutral medium is similar to the process occurring in the clumpy tori of AGN.
Therefore, we choose to fit the continuum of GX 301-2 with the ``clumpy torus''
(Ctorus\footnote{https://heasarc.gsfc.nasa.gov/xanadu/xspec/models/Ctorus.html}) model of \citet{Liu14}, which is implemented as a table model in {\sc isis}. 
Ctorus calculates the spectrum of scattered X-ray emission emerging from a clumpy torus, given an input power-law spectrum with a photon index $\Gamma$ between 1.5 and 2.5. 
The model includes the absorbed intrinsic continuum (transmitted component) and the scattered component. The
normalization of the scattered component is determined by the normalization of the input power law ($K_{\rm {pl}}$). The column densities seen by the transmitted and scattered components ($N^{\rm T}_{\rm H}$ and $N^{\rm R}_{\rm H}$, respectively; Table 2) are treated as independent fit parameters, because the absorbing column density along our
line of sight may be different from the average column density through the scattering material.
The clumpiness is represented by the average cloud number along any line
of sight ($N$, ranging between 2 and 10, the smaller the $N$, the more clumpy the gas) 
and the filling factor (assumed here to be 0.1). The densities and sizes of clumps are determined 
by $N^{\rm R}_{\rm H}$, $N$, and the filling factor \citep[see][for more details]{Liu14}.
We assumed an inclination angle of 90$^\circ$ (edge-on) to approximate a spherical geometry.

When fitting the spectra of GX 301-2, we found that the best-fitting photon indices always reach the lower limit of the model, $\Gamma = 1.5$. Therefore, we 
fixed the photon index to 1.5. To verify that this is a reasonable approximation, we extracted and 
analysed an archival {\it NuSTAR} spectrum 
taken at an orbital phase of $\approx0.9$ as mentioned in \S 2. 
We first fitted the spectrum between 15 and 50 keV using a power-law with a smoothed 
high-energy cutoff model \citep[eq. 2 in][]{LaB01}.
We found a photon index of $1.39\pm0.08$ with a cutoff energy of $23.9\pm0.4$ keV 
and a folding energy of $8.7\pm0.3$ keV (blue line in Figure 4). The reduced $\chi^2=1.0$.
Then we fitted the spectrum between 5 and 50 keV using an absorbed power-law with high-energy 
cutoff, plus a Gaussian
line around 6.4 keV, and found a photon index of $1.35\pm0.03$ with a cutoff energy of
$23.7\pm0.3$ keV, 
a folding energy of $8.6\pm0.2$ keV, a line with centroid of $6.37\pm0.01$ keV and $\sigma$ 
of $50^{+30}_{-50}$ eV, and a column density of $1.4\pm0.1\times10^{23}$ cm$^{-2}$
(red line in Figure 4). The reduced $\chi^2=1.1$. The residuals 
are mainly around 7.1 keV (Fe \Kb line). Therefore, a photon index 
of 1.5 is not far from those obtained from \nus data. We note that our main results 
are not affected by the
exact slope of the power-law used to approximate the continuum.

Ctorus model has also a fluorescent component, however, it does not include the Compton shoulder of
the Fe \Ka line currently. The rest-frame energies adopted for the fluorescent lines also have some 
discrepancies with
those measured in the laboratory \citep{Bea67}, which makes the usage of the fluorescent component 
not straight forward.
Therefore, we use Gaussian to model the emission lines.
As the narrow Fe \Ka line is very close to 6.4 keV, we use two 
Gaussian (Table 2) with centroids at 6.404 keV and 6.391 keV, to 
represent the \Ka1 and \Ka2 doublet \citep{Bea67}.
The intensity ratio between \Ka1 and \Ka2 is set to 2:1, and their
velocity shifts and Gaussian widths are fixed to be the same for the two components. 
The neutral Fe \Kb line is also modelled as a
Gaussian (Table 2) centered at 7.058 keV.
We add another Gaussian with an energy around 6.3 keV to represent the Compton shoulder.
In addition, we find relatively weak line features around 6.7 keV, 6.9 keV, and 7.5 keV, which we interpret as He-like Fe, H-like Fe, and neutral Ni \Ka lines. We also modelled them as Gaussian (Table 2) with a fixed width of 1 eV.

First, we fitted the spectrum from interval A (Table 2 and black data points in the top panel of Figure 5). We found that the direct power-law component contributes only $\approx$1/3 of the observed continuum emission around 6.5--9 keV; the rest comes from the scattered continuum. If both the Fe \Ka line and the continuum are dominated by photons scattered in Compton-thick gas, this could explain the similarity of their phase profiles; however, the location and geometry of the scattering medium must be such as to preserve the pulsation signal. 

For interval B, first we tried freezing the power-law normalization and the scattering column density in Ctorus to the
best-fitting values found for interval A, and only allowed the absorbing column density as a free parameter. We wanted
to test the scenario that the increase in the observed continuum flux was solely due to a decrease in the absorbing
column density. With this assumption, we get a column density of $N^{\rm T}_{\rm H} = (8.9\pm0.1)\times10^{23}$
cm$^{-2}$ and $\chi^2_{\nu} =2.3$. The model does not do a good job of reproducing the continuum at the low- and high-energy ends.
We obtain a better fit ($\chi^2_{\nu} =1.26$; green data points in the top panel of Figure 5) when we thaw the
normalization of the intrinsic power-law and the reprocessing column density. The power-law normalization is about twice the value in interval A (Table 2); the transmitted component now accounts for about half of the observed continuum flux above 6.5 keV. If the increase of the intrinsic luminosity is real, and no other changes occur in the distribution of the scattering medium, we would expect to see a corresponding increase of the fluxes from the Fe lines; however, all line fluxes are found to be similar between intervals A and B, within the uncertainties (Table 2). What does change in the line properties is the loss of the pulsation signal. 

Finally, we fitted the same model to the spectra of intervals C (Figure 5, middle panel) and D (Figure 5, bottom
panel). The trend for these late intervals is that the normalization of the intrinsic power-law emission decreases, and the scattering column density also decreases.  
The ratio between the measured fluxes of the Compton shoulder and of the Fe \Ka line is about 35\% and 32\% for intervals A and B, decreasing to about 29\% and 17\% for intervals C and D, consistent with the reduced scattering column density. The combined result of both changes is that 
the contribution of the absorbed intrinsic emission is larger than half of the observed continuum flux above 6.5 keV for intervals C and D. The fluorescence line fluxes also decrease, either because of the reduction in the intrinsic continuum flux, or because of the decrease in the scattering optical depth, or both.

The measured offsets of the Fe \Ka line from those of totally neutral Fe are very small; 
it indicates that the Fe \Ka photons are mostly due to nearly neutral Fe ions. 
This is consistent with the relatively low temperature found by \citet{Wat03}.
The best-fitting Gaussian width $\sigma$ (defined as $\sigma$ = 1/2.35 of the full-width-half-maximum, FWHM) of the Fe
\Ka line is around 5--9 eV during the four observational intervals. These values are consistent with those
independently reported by \citet{TY18}, also based on \cha data. Instead, they are smaller than those reported in
previous studies \citep[e.g.][]{End02,Fur11}, as mentioned in Section 1. The discrepancy is most likely due to the
higher spectral resolution of \cha HEG. As the natural width of the Fe \Ka line is FWHM $\approx$ 3 eV, the Gaussian widths fitted in our spectral models are consistent with a velocity broadening ($\approx$1/2\,FWHM) of $\approx$200--400 km s$^{-1}$. 
The fitted values of $\sigma$ for the Fe \Kb line are a little higher than those of the 
Fe \Ka line; those measurements may be affected by the Fe absorption edge near 7.1 keV.
The inferred equivalent widths (EW) of the Fe \Ka line and the absorption-corrected intrinsic power-law luminosities for
the four intervals are also listed in Table 2. The smaller EW of interval B is due to the larger
continuum level of interval B, compared with interval A. 

The fitted $N^{\rm T}_{\rm H}$ and $N^{\rm R}_{\rm H}$ are similar and decrease from interval A to D.
The fitted averaged cloud number increases from interval A to D. It seems that from interval A to D,
the medium is changing from a clumpy dense region to a less-clumpy less-dense region. We note that
the Ctorus model used has some deficiencies, for example, the input powerlaw 
does not include a high energy cutoff, as is required for GX 301-2; the photon index is limited to
1.5-2.5; the torus geometry may not represent the true cases.
Further model developements are needed to obtain more conclusive results. 

\section{Discussion and Conclusions}

The key new result of this paper is that we detected a pulsating Fe \Ka line with the same period as the continuum, during the first 7 ks (interval A) of a \cha observation of GX 301-2. The phase profile of the Fe \Ka line was also similar to that of the continuum. The pulsation signal of the Fe \Ka line then disappeared; at that time, the continuum flux increased by a factor of 3 and kept its pulsation profile; the line flux increased only by about 10\%. 

X-ray fluorescence lines are one of the hallmarks of neutron star HMXBs \citep[{\it e.g.},][]{LC93,Bas78,HM77}; they may be formed in many locations: in the accretion column, on the surface of the donor star or of an accretion disk, in the stellar wind, or in an accretion stream, shell or wake. Usually, the line emission is not pulsed \citep[{\it e.g.},][]{Ber15,Suc12,Wil11,Lei09,Pau02}.
A few well-studied exceptions, where fluorescent line emission from iron, oxygen and/or neon is pulsed, are 4U\,1626$-$67 \citep{Ber15,Sch01,Ang95}, Hercules X-1 \citep{Vas13,Zan04,Cho94}, and GX 1+4 \citep{Yos17a,Yos17b,Kot99}.

GX 301-2 was one of the X-ray pulsars in which the pulsed fraction of the Fe \Ka line was always absent or much smaller than the pulsed signal in other energy bands; this was explained as the effect of smearing of the pulsed signal in an isotropic line-emitting region around the neutron star \citep{Tas91,End02,Fur11}. Instead, we have now shown that at some epochs, the pulsed fraction of the line is as high as that of the continuum. Thus, the simplest qualitative explanation of this finding is that during those epochs, the pulsating Fe \Ka line originates in an anisotropic gas structure, illuminated by the pulsed emission of the neutron star. More remarkably, we have shown that the line pulsation disappears at a phase $\phi = 0.000 \pm 0.003$ (extrapolating from the ephemeris and period derivative solution of \citealt{Dor10}).

The inferred column density of the scattering medium responsible for the \Ka
fluorescence lines (inferred from the relative strength of the \Ka Compton shoulder), and 
for the scattered continuum (inferred from broad-band spectral modelling) are similar, that is
$\approx1.5\times10^{24}$ cm$^{-2}$. 
Such a high column density cannot be produced by the quasi-isotropic stellar wind of a super-giant donor, but can be
reached in the transient accretion stream formed between the donor star and the neutron star near periastron
\citep{Ste88}. Evidence for the existence of this dense stream flow comes from studies of optical lines \citep{Kap06},
and the  recurrent pre-periastron X-ray flare \citep[][and references therein]{LK08}. 
When the neutron star enters and leaves the accretion stream, an anisotropic configuration will 
be formed.
Specifically, the {\it Chandra} observation studied in this work was taken shortly after the flare; 
in the accretion stream scenario, it might correspond to the phase when the neutron star 
leaves the accretion stream. 
The illumination of the anisotropic stream distribution would 
produce the pulsating Fe \Ka line. When the neutron star moves further away from the stream, 
the contribution of the pulsating Fe \Ka line becomes less important.
Such a scenario also predicts a pulsating Fe \Ka line
when the neutron star enters the accretion stream, i.e., before the X-ray flare. 
Moreover, the pulsating behaviour should repeat at similar phases every orbital cycle.
Such a scenario can be tested with future observations.

Another possible anisotropic structure is the surface of the donor star. At periastron, 
the donor star subtends a solid angle $\Omega_{\rm p} \approx 1.88$ sr, that is, it is seen by 
the neutron star with an apex half-angle $\theta\approx 45^{\circ}$. Thus, for a moderate 
misalignment between the beam of X-ray emission and the normal to the
orbital plane, the beamed X-ray emission of the neutron star could directly illuminate 
the surface of the donor star, and produce the pulsating Fe \Ka line. As the neutron star 
moves away from periastron, the solid angle subtended by the donor star decreases. 
At apastron, the subtended solid angle is only $\Omega_{\rm a} \approx 0.22$ sr,
corresponding to a half-angle $\theta \approx 15^{\circ}$. In this scenario, we suggest 
that the disappearance of the pulsating line signal corresponds to the X-ray beam missing 
the stellar surface. 
Such scenario predicts that the pulsating phenomenon repeats periodically at periastron, 
and can also be tested with future observations.

The phase profile of the Fe \Ka line of GX 301-2 during interval A is quite different from 
those reported in the literature for other sources with pulsating line emission.
 For example, the phase profile of the Fe \Ka line from Her X-1 shows a sharp minimum
around the continuum peak, which might indicate a hollow cone of 
the accretion structure \citep{Vas13}. The \OVII\ line fluxes of 4U 1626$-$67 show a variation stronger
than that of the continuum, which could be due to variable illumination of the warped region of 
an accretion disk \citep{Ber15}. In contrast, the phase profile of the Fe \Ka line of GX 301-2
during interval A is similar to that of the continuum. It indicates that a major part of the 
observed continuum might come from the Compton-thick medium producing the Fe \Ka line, as evidenced
by the spectral modelling in \S 3.2.

In principle, even a homogeneous, spherically symmetric scattering medium illuminated by a central X-ray pulsar can produce a fluorescent signal that appears pulsed to the distant observer, because of the finite-light-speed effect \citep{Yos17b}. However, the characteristic size of the medium for this effect to be significant is of order of the pulsar spin period times the speed of light. In the case of GX 301-2, that would require a scattering medium with a size $\sim$10$^{13}$ cm, implausibly larger than the binary system itself.

The possibility that the observed continuum could have a significant contribution from 
reprocessed emission may change the interpretation of the time delays between the line and continuum.
Using \xmm data, \cite{Fur11}
found no time delays between Fe \Ka line and continuum in the time range of 2--5,000 s, and they inferred
a fluorescent region
smaller than 2 lt-s ($6\times 10^{10}$ cm). 
However, if a major part of observed continuum come from scattering, far from the neutron star surface,
the lack of relative time delays would no longer constrain the size of the region. 
A reanalysis of \xmm data is needed to check this possibility. 

Another way to constrain the region of the fluorescent lines is through their velocity broadening. 
Thanks to the \cha HEG resolution, we have shown that typical velocity broadening of 
the Fe \Ka line is $\approx$200--400 km s$^{-1}$, a few times lower than previously inferred. 
This is comparable to the typical terminal velocity of the stellar wind and the 
ballistic motion of an accretion stream. If the broadening is due to gas in virial motion, 
such velocities 
would correspond to distances $\sim5\times10^{10}$--$2\times10^{11}$ cm for a neutron star of 
1.5 $M_\odot$, similar to the accretion radius of GX 301-2 ($\sim10^{11}$ cm). Such a relatively 
large scale is consistent with fluorescence lines from cool and nearly neutral Fe ions. 
We note that it is 
much larger than the magnetosphere of a neutron star even with a field of $10^{14}$ G
($4\times10^{9}$ cm). 
On the other hand, 
the fluorescent region can not be larger than the light travel distance of 
a significant fraction of the pulse period (e.g., 100 ls), otherwise the pulsation signal would be
smeared out.   
In principle, any temporary anisotropic gas structure on scales of $\sim10^{11}$ cm with enough covering 
factor, like a torus or a warped accretion disk, could be invoked to explain the pulsating
Fe \Ka line. Such gas structures are likely caused by the dynamical interaction between the 
donor star and the neutron star near periastron. Detailed hydrodynamical simulations are needed 
to test the existence of such structures.

The increase in the observed continuum after the first 7 ks of the observation, accompanied by the non-increase of the
line emission and the disappearance of the pulsed line signal remain puzzling, and can be due to a combination of
increased intrinsic emission, decreased absorption column density along our line of sight, and decreased column
density of the scattering medium. Simultaneous observations in wide X-ray bands are needed to disentangle these 
effects. Nonetheless, using only the \cha HEG data, we have already shown a decreasing trend in the column density of
the scattering medium in the second half of the observation. This is seen both from our broad-band continuum
modelling, and from the progressively lower flux ratio between the Compton shoulder and the Fe \Ka line.

\section*{Acknowledgements}
We thank our anonymous referee for valuable suggestions that improve the work significantly, 
and Youjun Lu for helpful discussions. JL is supported by NSFC grant 11203032.
This research is based on data obtained from the \cha Data Archive.
We made use of software obtained from the High Energy Astrophysics Science Archive Research Center (HEASARC), a service of the Astrophysics Science Division at NASA/GSFC and of the Smithsonian Astrophysical Observatory's High Energy Astrophysics Division.

\end{document}